\documentclass[11pt]{article}
\usepackage{latexsym}
\usepackage{amssymb}
\usepackage{amsmath}
\usepackage[hypertex]{hyperref}
\usepackage{graphicx}% Include figure files

\begin{document}

\begin{flushright}
SLAC-PUB-11953\\
hep-ph/0607090 \\
\end{flushright}

\vspace{2.5cm}

\begin{center}

{\bf\LARGE Gravitational Gauge Mediation} \\

\vspace*{1.5cm}
{\large Ryuichiro Kitano} \\
\vspace*{0.5cm}

{\it Stanford Linear Accelerator Center, Stanford University,
                Stanford, CA 94309} \\

\vspace*{0.5cm}

\end{center}

\vspace*{1.0cm}

\begin{abstract}
It is often the case that the naive introduction of the messenger
sector to supersymmetry breaking models causes restoration of
supersymmetry. We discuss a possibility of stabilizing the
supersymmetry breaking vacuum by gravitational interaction.

\end{abstract} 

%%%%%%%%%%%%%%%%%%%%%%%%%%%%%%%%%%%%%%%%%%%%%%%%%%%%%%%%%%%%%%%%%%%%%%%%%%%%
\newpage
\baselineskip 18pt

\section{Introduction}

In scenarios where the electroweak scale is stabilized by
supersymmetry, it is expected that future collider experiments give us
hints of how the standard model sector feels the supersymmetry
breaking. It is now important to discuss possible mechanisms for
supersymmetry breaking and its mediation.

In field theory, spontaneous breaking of supersymmetry is quite easily
realized. An obvious example is the Polonyi model in which a singlet
chiral superfield $S$ has a linear superpotential $W = \mu^2 S$. This
model also serves as the low energy effective theory of the
O'Raifeartaigh model by introducing a non-minimal K{\" a}hler
potential term $K \ni - (S^\dagger S)^2/\widetilde{\Lambda}^2$ with
$\widetilde{\Lambda}$ the mass scale of the particles which are
integrated out.
The same effective theory can also be realized in models with strongly
coupled gauge theories such as in the IYIT model in
Ref.~\cite{Izawa:1996pk,Intriligator:1996pu}.
Recently it has been noticed that a wide class of supersymmetric QCD
leads to the above effective theory around the meta-stable
vacuum~\cite{Intriligator:2006dd}.

Even though spontaneous supersymmetry breaking may be a common feature
of quantum field theory, mediation of the supersymmetry breaking to
the standard model sector is not so simple.
Of course, gravity mediation is the simplest way to communicate with the
hidden sector. However, we study the possibility of gauge mediation
partly because it provides us with a solution to the flavor
problem~\cite{Dine:1993yw} (See \cite{Dine:1981za} for earlier works.).

A naive way of realizing gauge mediation is to introduce vector-like
messenger particles $q$ and $\bar{q}$ which carry standard model
quantum numbers and assume a coupling $W \ni -\lambda S q \bar{q}$ in
the superpotential.
Assuming non-vanishing vacuum expectation values in the lowest and the
$F$-component of $S$, the gaugino masses are obtained by the formula
$m_{1/2} = (\alpha/4\pi)(F_S/S)$ at one-loop level.
However, this model has a supersymmetric and hence global minimum where
$S=0$ and $q = \bar{q} = \sqrt{\mu^2 / \lambda}$.
Therefore the question is whether it is possible to have a meta-stable
vacuum with non-vanishing value of $S$~\cite{Arkani-Hamed:1996xm}. For
example, in the O'Raifeartaigh model, the sign of the coefficient of
the K{\" a}hler term $(S^\dagger S)^2/\widetilde{\Lambda}^2$ is
negative, which stabilizes the field $S$ at the origin. Therefore,
there is no meta-stable vacuum away from the origin.

Mechanisms of realizing a (local) minimum away from the origin have
been discussed in the literature. In Ref.~\cite{Murayama:1997pb}, it
is shown that the inverted hierarchy mechanism can produce a local
minimum at a very large value of $S$ if $S$ is non-singlet under some
gauge interaction. The possibility of having a (local) minimum via a
non-perturbative effect has also been discussed in
Ref.~\cite{Izawa:1997gs} (See also
\cite{Intriligator:1996pu,Shirman:1996jx,Arkani-Hamed:1997ut,Chacko:1998si}
for discussions of the vacuum structure of the IYIT model.).
The introduction of a bare mass term for the messenger field also
makes the vacuum meta-stable as discussed in Ref.~\cite{Izawa:1997gs,
Izawa:2005yf}.
In the context of supersymmetry breaking in chiral gauge theories, it
has been discussed that a runaway direction which is lifted only by
non-renormalizable operators has a minimum at a very large field
value~\cite{Poppitz:1996fw}.

In this note, we argue that the situation of supersymmetry restoration
by the messenger particles can be cured once we include the
supergravity effect even if the origin is the unique minimum in the
limit of the Planck scale $M_{\rm Pl}$ to infinity.
Although the gravity effect is always suppressed by $1/M_{\rm Pl}$, it
can be large enough to stabilize $S$ away from the origin ($S \sim
\widetilde{\Lambda}^2/M_{\rm Pl}$) when $\widetilde{\Lambda} \gtrsim
10^{13}$~GeV.

\section{Meta-stable vacua in supergravity}

The model we will analyze is the following:
\begin{eqnarray}
K = S^\dagger S - \frac{(S^\dagger S)^2 }{\widetilde{\Lambda}^2}
+ q^\dagger q + \bar{q}^\dagger \bar{q} \ ,
\label{eq:kahler}
\end{eqnarray}
\begin{eqnarray}
 W = \mu^2 S - \lambda S q \bar{q} + c\ .
\label{eq:super}
\end{eqnarray}
The chiral superfield $S$ is a singlet field, $q$ and $\bar{q}$ are
the messenger fields which carry standard model quantum numbers, and
$\lambda$ is a coupling constant.
The constant term $c$ does not have any effect if we neglect gravity
interactions, but it is important for the cancellation of the
cosmological constant. If we neglect the constant term $c$, the
Lagrangian possesses an $R$-symmetry with the charge assignments
$R(S)=2$, $R(q) = R(\bar{q}) = 0$.
We first discuss the model without gravity effect ($M_{\rm Pl} \to
\infty$ limit).

For $\lambda = 0$, this model breaks supersymmetry and $S$ is stabilized
at the origin, $S=0$, by the supersymmetry breaking effect.
However, by turning on the $\lambda$ coupling, $q$ and $\bar{q}$
acquire tachyonic mass terms ($B$-term) which make the vacuum
unstable. The true minimum is at $S=0$ and $q = \bar{q} =
\sqrt{\mu^2/\lambda}$ where supersymmetry is unbroken. Therefore,
there is no supersymmetry breaking vacuum in this model.

We assumed here that the sign of the second term in the K{\" a}hler
potential is negative as it is the case in the O'Raifeartaigh model.
In general, the sign can be positive, and sometimes it is even
uncalculable if this term is originated from some strong dynamics. When
it is positive, the $S$ field may be stabilized away from the origin and
the $q$ and $\bar{q}$ directions are also stabilized at $q = \bar{q} =
0$ by the supersymmetric mass terms.
However, we consider the case with a negative sign where the origin is
a stable minimum.\footnote{
In Ref.~\cite{Chacko:1998si}, the presence of a minimum at the
origin is shown in the IYIT model.}
As we will see later, the supersymmetry breaking minimum reappears
when we turn on the gravity effect even in that case.

First, we need to estimate the perturbative quantum correction to the
K{\" a}hler potential for $S$ coming from the interaction term
$\lambda S q \bar{q}$, which may be more important than the gravity
effect.
We can explicitly calculate the term at one-loop level:
\begin{eqnarray}
 K_{\rm {1 \mbox - loop}} = - \frac{\lambda^2 N_q}{(4 \pi)^2} S^\dagger S
\log \frac{S^\dagger S}{\Lambda^2}\ ,
\end{eqnarray}
where $N_q$ is the number of components in $q$ and $\bar{q}$. For
example, $N_q = 5$ if $q$ and $\bar{q}$ transforms as ${\bf 5}$ and ${\bf
\bar{5}}$ under SU(5)$_{\rm GUT}$.
Higher order perturbative contributions, including the dependence on
the artificial scale $\Lambda$, will be minimized by taking the
running coupling constant $\lambda$ to be the value near the scale
$S$.
This term also tends to make $S=0$ stable.

Now we include the gravity effect in this model.
The scalar potential of the supergravity Lagrangian is given by
\begin{eqnarray}
 V = e^G (G_S G_{S^\dagger} G^{S S^\dagger} + G_q G_{q^\dagger} 
+ G_{\bar{q}} G_{\bar{q}^\dagger}  - 3 ) + \frac{1}{2} D^2\ ,
\end{eqnarray}
where $G \equiv K + \log |W|^2$ and we set $M_{\rm Pl} = 1$. $G_X$ is
the derivative of $G$ with respect to the field $X$, and $G^{S
S^\dagger}$ is the inverse of the K{\" a}hler metric. $D^2/2$
represents the $D$-term contributions.
We can easily find the supersymmetric minimum, that is a solution of the
equations:
\begin{eqnarray}
 G_S = G_q = G_{\bar{q}} = 0, \ \ \ q = \bar{q}\ .
\end{eqnarray}
The solution is
\begin{eqnarray}
 q = \bar{q} = \sqrt{\frac{\mu^2}{\lambda}} 
+ O\left( \frac{c}{\lambda M_{\rm Pl}^2} \right)\ ,\ \ \ 
 S = O\left( \frac{c}{\lambda M_{\rm Pl}^2} \right)\ .
\end{eqnarray}
The gravity effect is a slight shift of the values of order $c/(\lambda
M_{\rm Pl}^2)$
which is $O(\mu^2/ (\lambda M_{\rm Pl}))$ if we assume the cancellation of the
cosmological constant at the meta-stable supersymmetry breaking vacuum
below.

Another minimum can be found with the assumption of $q = \bar{q} =
0$. The potential is simplified to
\begin{eqnarray}
 V = e^G (G_S G_{S^\dagger} G^{S S^\dagger}  - 3 ) \ .
\end{eqnarray}
The equation $V_S = 0$ with the phenomenological requirement $V=0$,
cancellation of the cosmological constant, leads to
\begin{eqnarray}
 (G^{S S^\dagger})_{,S}
= - (G^{S S^\dagger})^2 \left[
- \frac{\kappa}{S} - \frac{4 S^\dagger }{\widetilde{\Lambda}^2}
\right]
\simeq \frac{2 \sqrt{3}}{3 M_{\rm Pl}}\ ,
\end{eqnarray}
where $\kappa = \lambda^2 N_q / ( 4 \pi)^2$. For $\kappa \lesssim
(\widetilde{\Lambda} / M_{\rm Pl} )^2$, the minimum is at 
\begin{eqnarray}
 S \simeq \frac{\sqrt{3} \widetilde{\Lambda}^2}{6 M_{\rm Pl}}\ .
\end{eqnarray}
Supersymmetry is broken there with $F_S \simeq \mu^2$.
By taking the limit $M_{\rm Pl} \to \infty$, this minimum moves to $S
\to 0$ and the meta-stable vacuum disappears. However, with a finite
value of $M_{\rm Pl}$, the supersymmetry breaking and supersymmetric
vacua are at the different places.

In the $\lambda \to 0$ limit, there is no supersymmetric vacuum as in
the case without gravity, but the minimum is not at the origin ($S\sim
\widetilde{\Lambda}^2/M_{\rm Pl}$), even though the sign of the
$(S^\dagger S)^2$ term in K{\" a}her potential is negative. In the
supergravity Lagrangian, the origin is no longer a symmetry enhanced
point because the $R$-symmetry is explicitly broken by the
$c$-term. By turning on the $\lambda$ coupling, the supersymmetric
vacuum appears near the origin, but it is separated from the
supersymmetry breaking minimum. The disappearance of the supersymmetry
breaking minimum by the finite $\lambda$ coupling seen before was an
artifact of the approximation $M_{\rm Pl} \to \infty$.

The stability of the vacuum can be checked by looking at the mass matrices
of the $S$, $q$, and $\bar{q}$ fields. The matrices are given by
\begin{eqnarray}
 m_{S}^2 \simeq \frac{\mu^4}{\widetilde{\Lambda}^2} \cdot
\left(
\begin{array}{cc}
 4 & - 6 \kappa ( M_{\rm Pl}/ \widetilde{\Lambda} )^2 \\
- 6 \kappa ( M_{\rm Pl}/ \widetilde{\Lambda} )^2 & 4\\
\end{array}
\right)\ ,\ \ \ 
 m_{q}^2 \simeq
\left(
\begin{array}{cc}
 \lambda^2 \widetilde{\Lambda}^4 / (12 M_{\rm Pl}^2) & - \lambda \mu^2 \\
 - \lambda \mu^2 & \lambda^2 \widetilde{\Lambda}^4 / (12 M_{\rm Pl}^2)  \\
\end{array}
\right)\ .
\end{eqnarray}
Therefore, there is a stable minimum when $\kappa \lesssim
(\widetilde{\Lambda} / M_{\rm Pl} )^2$ and $\widetilde{\Lambda}^4 \gtrsim
\mu^2 M_{\rm Pl}^2 / \lambda $.

\begin{figure}[t]
 \begin{center}
  \includegraphics[height=6.5cm]{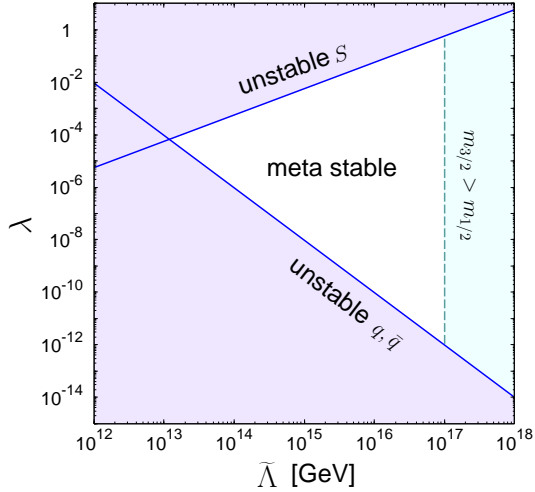}
\end{center}
\caption{The region of the parameter space where the supersymmetry
 broken minimum is meta-stable.}
 \label{fig:region}
\end{figure}

There is another phenomenological requirement that the gauge mediation
effect, the gaugino masses, is of order 100~GeV. This fixes the
relation between the parameters $\mu^2$ and $\widetilde{\Lambda}$ as
follows:
\begin{eqnarray}
 \mu^2 \simeq \left( \frac{\alpha}{4 \pi} \right)^{-1} \frac{M_W
  \widetilde{\Lambda}^2}{M_{\rm Pl}}\ ,
\label{eq:relation}
\end{eqnarray}
where $M_W$ is the electroweak scale.  With this relation, we show in
Fig.~{\ref{fig:region}} the parameter region where the supersymmetry
breaking vacuum is meta-stable.
A gravitational stabilization of the vacuum is possible for
$\widetilde{\Lambda} \gtrsim 10^{13}$~GeV. For large $\lambda$, the
one-loop correction to the $S$ potential destabilizes the vacuum, and
too small $\lambda$ leads to unstable $q$ and $\bar{q}$ direction
because of the small supersymmetric mass terms.
The tunneling rate into the supersymmetric vacuum is small enough for
$\widetilde{\Lambda} \gtrsim 10^{11}$~GeV, which is satisfied in whole
the viable region~\cite{Coleman:1977py}.
For $\widetilde{\Lambda} \gtrsim 10^{17}$~GeV, the possible gravity mediation
effect on the gaugino masses from the operator $SW^\alpha W_\alpha /
M_{\rm Pl}$ is larger than the gauge mediation. Note that this term is
forbidden if we impose the approximate $R$-symmetry discussed before.

The mass of the $S$ field depends on $\widetilde{\Lambda}$. With the
relation in Eq.~(\ref{eq:relation}), we obtain
\begin{eqnarray}
 m_S \simeq 100~{\rm GeV} \left(
\frac{\widetilde{\Lambda}}{10^{16}~{\rm GeV}}
\right)\ .
\end{eqnarray}
The gravitino mass $m_{3/2}$ is 
\begin{eqnarray}
 m_{3/2} \simeq 1~{\rm GeV} \left(
\frac{\widetilde{\Lambda}}{10^{16}~{\rm GeV}}
\right)^2\ .
\end{eqnarray}

\section{Ultraviolet completion}

There are several possibilities for the underlying microscopic models
which give the effective theory defined in Eqs.~(\ref{eq:kahler}) and
(\ref{eq:super}). An obvious example is the O'Raifeartaigh model as
discussed before. The mass scale of the fields which are integrated out
is identified with $\widetilde{\Lambda}$.

There is another interesting possibility that the scale
$\widetilde{\Lambda}$ is identified with the dynamical scale of the
strongly coupled gauge theory and the linear term $\mu^2 S$ originates
from the mass term of the quarks in that theory. This possibility is
realized quite simply in the models of
Ref.~\cite{Intriligator:2006dd}. In the SO($N_c$) gauge theory with $N_c
- 4$ flavor quarks, there is a branch where the quarks confine and there
is no non-perturbatively generated superpotential. If there is a mass
term for the quarks $T$ in the $N_c$ dimensional representation, $W = m
T^2$, the low energy effective theory is
\begin{eqnarray}
 W_{\rm eff} = m M\ ,
\end{eqnarray}
where $M$ is the meson field $M \sim T^2$. Therefore there is no
supersymmetric vacuum in this branch. By assuming the presence of the
coupling of the messenger fields $q$ and $\bar{q}$ to the operator
$T^2$, the effective superpotential is identical to that in
Eq.~(\ref{eq:super}) with $\mu^2 \sim m \widetilde{\Lambda}$ and $S
\sim M /\widetilde{\Lambda}$.  If the coupling is suppressed by the
Planck scale, i.e., $W \ni T^2 q \bar{q} / M_{\rm Pl}$, the $\lambda$
parameter is of order $\widetilde{\Lambda}/M_{\rm Pl}$, and the upper
limit on the $\lambda$ coupling, $\kappa \lesssim (\widetilde{\Lambda}
/ M_{\rm Pl} )^2$, is always satisfied.
The stability of the vacuum is ensured by the fact that the potential
grows for large $S$ at the classical level by the mass term of $T$,
and the classical analysis is reliable for $S \gtrsim
\widetilde{\Lambda}$.
If this stabilization is due to the $(S^\dagger S)^2$ term in the K{\"
a}hler potential, the K{\" a}hler potential in Eq.~(\ref{eq:kahler}) is
obtained.
For other gauge groups such as SU($N_c$) and Sp($N_c$), it is suggested
that there are similar vacua in models with $N_c$ and $N_c + 1$ flavors,
respectively.

For other numbers of flavors, it is shown that there are meta-stable
supersymmetry breaking vacua, for example, in SU($N_c$) with $N_c + 1
\leq N_f < 3 N_c / 2$~\cite{Intriligator:2006dd}. However, in those
cases, the relation between the dynamical scale $\Lambda$ and the
parameters in Eqs.~(\ref{eq:kahler}) and (\ref{eq:super}),
($\widetilde{\Lambda}$, $\mu^2$), is $\widetilde{\Lambda} \sim \sqrt{m
\Lambda}$ and $\mu^2 \sim m \Lambda$, respectively.
With this relation, $\mu^2 \sim \widetilde{\Lambda}^2$, we cannot
satisfy the relation in Eq.~(\ref{eq:relation}).\footnote{Strictly
speaking, the two scales $\mu^2$ and $\widetilde{\Lambda}^2$ can be
separated by assuming a mass hierarchy among the quarks. In that case,
$\mu^2 \sim m_L \Lambda$ and $\widetilde{\Lambda}^2 \sim m_H \Lambda$,
where $m_L$ and $m_H$ are masses of the light and heavy quarks,
respectively.  For $m_H \gtrsim \Lambda$, the discussion reduces to
the case with fewer flavors.  }
Exceptions are SO($N_c$) with $N_c - 3$ and $N_c - 2$ flavor theories
where there is no non-perturbatively generated superpotential. In the
$N_c - 3$ flavor model, there is a branch where low energy effective
theory has superpotential~\cite{Intriligator:1995id}:
\begin{eqnarray}
 W = f(t) S a^2 + \mu^2 S\ .
\end{eqnarray}
The chiral superfield $a$ consists of $N_f$ gauge singlet fields, and
$f(t)$ is an unknown function of $t = (\det S) (S a^2)$ with $f(0)
\neq 0$. We expect that the coupling constant between $S$ and $a^2$ is
$O(1)$ at low energy. In that case, the vacuum with $a = 0$ may be
meta-stable for $\widetilde{\Lambda} \gtrsim 10^{16}$~GeV according to
Fig.~\ref{fig:region}. In the $N_c -2 $ flavor model, the low energy
effective theory is U(1) gauge theory with a similar superpotential:
\begin{eqnarray}
 W = f(t) S a^+ a^- + \mu^2 S\ ,
\end{eqnarray}
where $a^+$ and $a^-$ are the monopoles and $t = \det S$.
The same conclusion applies in this case.

For these $N_f = N_c - 3$ and $N_f = N_c - 2$ models, it may be
possible that the fields $a$ or $a^{\pm}$ are actually the messenger
fields $q$ and $\bar{q}$ by gauging the subgroup of the flavor
symmetry SU($N_f$) and identifying it with the standard model gauge
group. However, since the meson field $S$ is a symmetric $N_f \times
N_f$ matrix which is stabilized only by the supersymmetry breaking
effect, i.e., $m_S \simeq \mu^2 / \widetilde{\Lambda} \lesssim
10$~TeV, it gives too large contributions to the beta function of the
standard model gauge couplings. A larger structure, e.g., introduction
of the partner of the unwanted light fields, is necessary for such a
scenario to be viable.

A particularly interesting scale for $\widetilde{\Lambda}$ is the GUT
scale $M_{\rm GUT}\sim 10^{16}$~GeV. From Eq.~(\ref{eq:relation}) and
$\mu^2 \sim m \widetilde{\Lambda}$, the mass parameter $m$ is $O(M_W)$
for $\widetilde{\Lambda} \sim M_{\rm GUT}$.
In this case, the parameter $m$ can be related to the $\mu$-parameter
in the minimal supersymmetric standard model, which is the only
explicit mass parameter in the model.\footnote{ The $\mu$-parameter
shouldn't be confused with the $\mu^2$ term in Eq.~(\ref{eq:super}).
For $\widetilde{\Lambda} \sim M_{\rm GUT}$, $\mu^2$ is an intermediate
scale such as $\mu^2 \sim (10^9~{\rm GeV})^2$, whereas the
$\mu$-parameter is always $O(100~{\rm GeV})$.  }
Indeed, the parameter $m$ can really be the $\mu$-parameter in the
scenario where the Higgs fields are composite particles of the strong
dynamics which we are discussing~\cite{Kitano:2006wm}. Moreover, the
same dynamics can be responsible for the dynamical breaking of the gauge
symmetry in grand unified theories as shown in
Ref.~\cite{Kitano:2006wm}.
In the conventional picture, the electroweak scale appears as a
consequence of the supersymmetry breaking, and there was a puzzle that
the supersymmetric parameter, the $\mu$-parameter, must be the same
size as the supersymmetry breaking parameters. This puzzle was
particularly sharp in the scenario of gauge mediation. However, in
this scenario, the electroweak scale is the scale which drives the
supersymmetry breaking and therefore there is no coincidence problem.
There are many possibilities for the origin of the scale $O(100~{\rm
GeV})$ such as the dynamical scale of another strongly coupled gauge
theory. A more attractive possibility of relating it to the cosmological
constant term, $c$-term in Eq.~(\ref{eq:super}), is pointed out in
Ref.~\cite{Kitano:2006wm}.

The IYIT model also gives the same effective theory in
Eqs.~(\ref{eq:kahler}) and (\ref{eq:super}).
The model is an Sp($N_c$) gauge theory with $N_f = N_c + 1$ flavors
with the superpotential:
\begin{eqnarray}
 W = y S Q Q\ ,
\end{eqnarray}
where $S$ is a singlet field and an anti-symmetric $2N_f \times 2N_f$
matrix, $Q$ is the quark in the $2N_c$ dimensional representation, and
$y$ is the coupling constant. By the quantum modified constraint, the
effective superpotential is
\begin{eqnarray}
 W = y \Lambda^2 S\ ,
\end{eqnarray}
with $\Lambda$ being the dynamical scale. Therefore $\mu^2 = y
\Lambda^2$. Near the origin of $S$, the correction to the effective K{\"
a}hler potential is calculated in Ref.~\cite{Chacko:1998si} to be
\begin{eqnarray}
 \delta K \sim 
- \frac{y^2}{(4 \pi)^2} \frac{(S^\dagger S)^2}{\Lambda^2}\ ,
\label{eq:kahler-iyit}
\end{eqnarray}
which stabilize the origin of the potential. The relation
$\widetilde{\Lambda} \simeq 4 \pi \Lambda / y$ is obtained. In order to
satisfy the relation in Eq.~(\ref{eq:relation}), the coupling constant
is determined to be $y \simeq 10^{-4}$. With this small value of $y$,
the field $S$ is stabilized at a large value, $S \sim (4\pi)^2 \Lambda^2
/ ( y^2 M_{\rm Pl})$ which must be smaller than $\Lambda / y$ so that
the effective K{\" a}hler potential in Eq.~(\ref{eq:kahler-iyit}) is
reliable. This constraint gives an upper limit on $\Lambda$ to be
$\Lambda \lesssim 10^{12}$~GeV which translates into
$\widetilde{\Lambda} \lesssim M_{\rm Pl} / 10$.
This is consistent with the region in Fig.~\ref{fig:region}.

\section{Summary}

We considered a gravitational stabilization mechanism of the
supersymmetry breaking vacuum in a simple gauge mediation model. The
gravitational interaction splits the supersymmetry breaking and
supersymmetric vacua for large enough values of the ``cut-off'' scale
$\widetilde{\Lambda} \gtrsim 10^{13}$~GeV.

The low energy effective model we analyzed can arise from many
microscopic theories of supersymmetry breaking. Therefore the
mechanism we discussed is applicable to a wide class of models.
The model possesses $R$-symmetry. The $R$-symmetry is unbroken at the
origin of the field $S$. However, the explicit breaking of the
$R$-symmetry in supergravity (by the cosmological constant) shifts the
vacuum from the origin. There is no unwanted Goldstone mode associated
with the $R$-symmetry breaking since the symmetry is broken
explicitly~\cite{Banks:1993en}.

\section*{Acknowledgments}

I thank Carola Berger for reading the manuscript. This work was
supported by the U.S. Department of Energy under contract number
DE-AC02-76SF00515.

\end{document}